\documentclass[fleqn,twoside]{article}
\usepackage{espcrc2}
\usepackage{epsfig}

\usepackage{graphicx}
\usepackage[figuresright]{rotating}

\newcommand{\Z}{{\sf Z \!\!\! Z}}
\newcommand{\1}{{\sf 1 \!\!\!\! 1}}
\newcommand{\beq}{\begin{equation}}
\newcommand{\eeq}{\end{equation}}

\newcommand{\AmS}{{\protect\the\textfont2
  A\kern-.1667em\lower.5ex\hbox{M}\kern-.125emS}}

\title{The deconfinement phase transition in Yang-Mills theory with
  general Lie group $G$}

\author{K. Holland\address[UCSD]{Department of Physics, University of
          California at San Diego, La Jolla, CA 92093, U.S.A.},
        M. Pepe\address[Bern]{Institute for Theoretical Physics,
          Bern University, Sidlerstrasse 5, CH-3012 Bern,
          Switzerland.}\thanks{Speaker at the Conference.},
        U.-J. Wiese\addressmark[Bern] \thanks{On leave from MIT.\newline
Work supported by the DOE under the grant DOE-FG03-97ER40546 and by
the Schweizerischer Nationalfond.}.} 

\begin{document}

\begin{abstract}
We present numerical results for the deconfinement phase transition in $Sp(2)$ and
$Sp(3)$ Yang-Mills theories in $(2+1)$-D and $(3+1)$-D. We
then make a conjecture on the order of this phase transition in
Yang-Mills theories with general Lie groups $G = SU(N), SO(N),
Sp(N)$ and with exceptional  groups $G = G(2), F(4), E(6), E(7),E(8)$.  
\end{abstract}

\maketitle

\section{Introduction and Overview}

Let us consider Yang-Mills theory with gauge symmetry group $G$ on a
periodic lattice in $(d+1)$-D with the Wilson action. This theory is invariant
under the multiplication by an element of the center subgroup 
$C(G)$ of $G$ of all time-like links in a given time-slice.  This
global center symmetry is unbroken at low temperatures and
it gets spontaneously broken at the deconfinement phase
transition. The Polyakov loop $\Phi$ transforms non-trivially under
this symmetry and thus is an order parameter for the deconfinement transition. 

Integrating out the spatial degrees of freedom of the 
Yang-Mills theory, one can write down an effective action for
$\Phi$. It describes a scalar model with global 
symmetry $C(G)$ in $d$-D: the $C(G)$-symmetric confined phase of the gauge theory
corresponds to the disordered phase of the scalar model, while the 
$C(G)$-broken deconfined phase has its counterpart in the
ordered phase. Svetitsky and Yaffe \cite{Sve82} argued that the
interactions in the effective description are short ranged. Hence, if the
deconfinement phase transition in the Yang-Mills theory is
second order, approaching criticality, the details of the
complicated short-ranged effective scalar action become progressively
unimportant. Only the dimensionality $d$ and the symmetry $C(G)$ of the
scalar model are relevant. Thus, one can exploit the
universality of the critical behavior to use a simple scalar model to
obtain information about the much more complicated Yang-Mills theory.  
If the phase transition is first order the correlation length
does not diverge: there are no universal features and the $G$-symmetric
Yang-Mills theory in $(d+1)$-D does not share the critical
behaviour with a $C(G)$-symmetric $d$-D scalar model.  

Yang-Mills theory on the lattice is naturally formulated in terms
of group elements while in the continuum the fundamental
field is the gauge potential, living in the algebra. An algebra can
generate different groups, however it is natural to expect that
lattice Yang-Mills theories whose gauge groups correspond to the same
algebra have the same continuum limit\cite{deF03a}.
Hence, instead of
$SO(N)$ we consider its covering group $Spin(N)$.
Keeping this in mind, we look at the center subgroups $C(G)$ of
the various simple Lie groups $G$ 
\vskip-.6cm
\begin{eqnarray}
&&\hskip-.7cm C(SU(N))=\Z(N);\hskip.8cm C(Sp(N))=\Z(2) \\ 
&&\hskip-.7cm C(SO(N))\hskip-.1cm\rightarrow \hskip-.1cm 
C(Spin(N))= \hskip-.1cm
\left\{\begin{array}{l} 
\hskip-.2cm\Z(2); \;\;\, N \;\mbox{odd} \\ 
\hskip-.2cm\Z(2)^2 ;\; N=4k \\
\hskip-.2cm\Z(4) ; \;\;\, N\hskip-.1cm=\hskip-.1cm4k\hskip-.1cm+\hskip-.1cm2 
\end{array}\right.\\
&&\hskip-.7cm C(G(2))= C(F(4))=C(E(8))=\{\1\}\\
&&\hskip-.7cm C(E(6))=\Z(3);\hskip.9cm C(E(7))=\Z(2)
\end{eqnarray}
\vskip-.2cm
Many numerical simulations in $(2+1)$-D and $(3+1)$-D have been performed
for $SU(N)$ Yang-Mills theory in order to investigate the order of the
deconfinement transition and -- if it is second order -- to
check the validity of the Svetitsky-Yaffe conjecture. 
A recent study on this subject can be found in \cite{Luc03}. The
currently known results are:\\ 
$\bullet$ $(3+1)$-D: for $N=2$ -- consistent with the Svetitsky-Yaffe
conjecture -- the deconfinement transition is
second order, in the universality class of the $3$-D Ising model; for
$N=3,4,6,8$ the phase transition is first order with no universal features.\\
$\bullet$ $(2+1)$-D: fluctuations are stronger and the deconfinement
phase transition turns out to be second order for $N=2,3,4$. In
agreement with the Svetitsky-Yaffe conjecture, the $2$-D 
universality classes are, respectively, those of the Ising, 3-state
Potts, and Ashkin-Teller models.

The $SU(N)$ branch is not a good choice to study the relation
between the order of the deconfinement transition and the size
of the group. In fact, when increasing the size 
$N^2\hskip-.1 cm-\hskip-.1cm 1$, also
the center $\Z(N)$ changes. In order to disentangle these two features we
have considered $Sp(N)$ Yang-Mills theory. In this case, the
available universality class is fixed and the relevance of the
size of the group on the order of the deconfinement transition
can be directly addressed. 

\section{$Sp(N)$ Yang-Mills Theory}

The matrices $U\in SU(2N)$ with the property
\beq\label{Spdef}
U^* = J U J^+,\;\;\; \mbox{where}\;\;\; J= i \sigma_2 \otimes \1
\eeq
form the symplectic group $Sp(N)$; $\sigma_2$ is the imaginary Pauli matrix
and $\1$ is the $N\times N$ unit matrix. From the
definition (\ref{Spdef}), it follows that the Hermitean generators $H$
of $Sp(N)$ satisfy the constraint $H^* = -J H J^+ = J H J$. Then we can write
the following general forms for $U$ and $H$
\beq
\label{groupalgebra}
U =  \left(\begin{array}{cc} W & X \\ - X^* &  W^* \end{array}\right),
\;\;
H =  \left(\begin{array}{cc} A & B \\   B^* & -A^* \end{array}\right)
\eeq
where $A,B,C,D$ are $N\times N$ complex matrices satisfying:
$A=A^+$, $B=B^T$, $WW^+ + XX^+ = \1$ and $W X^T = X W^T$.
Counting the number of degrees of freedom, it follows that $Sp(N)$
has $N(2N+1)$ generators and its rank is $N$. For
the center elements -- since they are multiples of the unit matrix --
it holds that $X=0$ and $W=W^*$: hence $C(Sp(N))=\Z(2)$. 
We note that (\ref{Spdef}) implies that $Sp(N)$ is a pseudo-real
group with the special cases $Sp(1)=SU(2)$ and $Sp(2)=Spin(5)$.
The $Sp(N)$ Yang-Mills theory on the lattice can be formulated in the
usual way in terms of group-valued link variables. We have carried out
our simulations using the Wilson plaquette action.

\section{Numerical Results}
In the next two subsections we report on the results of numerical simulations 
in $Sp(2)$ and $Sp(3)$ Yang-Mills theories in $(2+1)$-D and $(3+1)$-D. 
\subsection{$Sp(2)$ Yang-Mills Theory}
\begin{figure}[h]
\vskip-1.3cm
\begin{center}
\epsfig{file=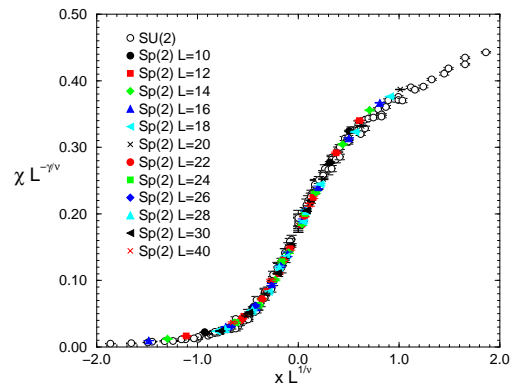,width=72mm,angle=0}
\vskip-1.5cm
\caption{\it Finite size scaling function of $\chi$;
$\nu$ and $\gamma$ are $2$-D Ising critical exponents; 
$x=(\beta/\beta_c - 1)$.}
\label{FSSchiSp2_2+1}
\end{center}
\vskip-1.1cm
\end{figure}

\begin{figure}[htb]
\vskip-1.6cm
\begin{center}
\epsfig{file=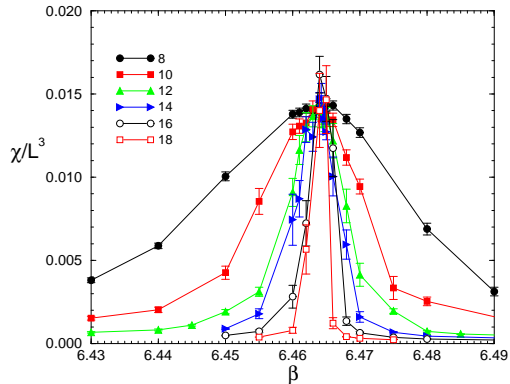,width=72mm,angle=0}
\vskip-1.5cm
\caption{\it Scaling of $\chi$ in $Sp(2)$ Yang-Mills theory on $L^3\times 2$
  lattices . We estimate $\beta_c=6.4643(3)$.}
\label{FSSchiSp2_3+1}
\end{center}
\vskip-1.5cm
\end{figure}

As a first step we have scanned the expectation value of the plaquette
from the strong to the weak coupling regime. We find no bulk phase
transition that might interfere with the study of the deconfinement
transition. In $(2+1)$-D we observe a second order deconfinement
transition, signalled by the broadening of the probability
distribution of $\Phi$ and, hence, by the increase of the Polyakov
loop susceptibility $\chi$ at criticality. A finite size scaling
analysis confirms the expectation that the universality class is that of
the $2$-D Ising model. Fig.~\ref{FSSchiSp2_2+1} shows the collapse of
$\chi$ data --~collected on lattices of different sizes $L^2\times 2$ and
at various gauge couplings $\beta$~-- on a single curve. 
We also plot rescaled $\chi$ data for $SU(2)$ Yang-Mills theory in
$(2+1)$-D: it again deconfines with a second order transition in the 2-D Ising universality class.
The two sets agree excellently.
In $(3+1)$-D --~contrary to what one might have expected~-- the probability
distribution of $\Phi$ in the critical region shows the coexistence of
the symmetric and of the broken phases, indicating that the
deconfinement transition is first order. Fig.~\ref{FSSchiSp2_3+1}
indeed shows that $\chi$ scales with the spatial volume $L^3$ at
criticality. 
\subsection{$Sp(3)$ Yang-Mills Theory}
\begin{figure}[htb]
\vskip-.4cm
\begin{center}
\epsfig{file=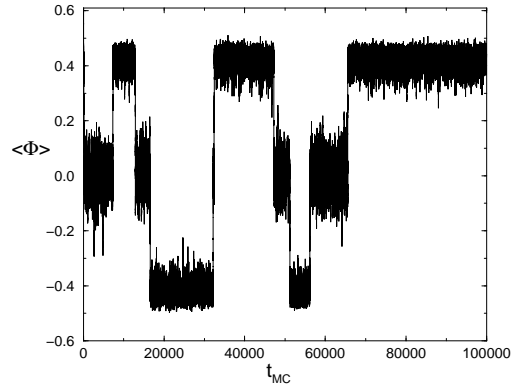,width=72mm,angle=0}
\vskip-1.2cm
\caption{$(3+1)$-D $Sp(3)$: \it tunneling events between the symmetric and the broken
phases.}
\label{PolyhistorySp3_3+1}
\end{center}
\vskip-1.3cm
\end{figure}

The results of $Sp(2)$ Yang-Mills theory show that in $(2+1)$-D
fluctuations are stronger than in $(3+1)$-D and the deconfinement
transition is second order. Expecting that the larger the group the
weaker the fluctuations, we have also investigated 
the deconfinement transition in $Sp(3)$ Yang-Mills theory. Consistent
with this picture, we find that $(2+1)$-D $Sp(3)$ Yang-Mills theory
has a first order deconfinement transition. The probability
distribution of $\Phi$ in the critical region, indeed displays the
coexistence of the broken and of the symmetric phases.  
In $(3+1)$-D -- similar to the $Sp(2)$ case -- $Sp(3)$ Yang-Mills
theory deconfines with a first order transition. 
Fig.~\ref{PolyhistorySp3_3+1} shows tunneling events between the
coexisting symmetric and broken phases as a function of Monte
Carlo time $t_{\mbox{\tiny {MC}}}$.   
$Sp(3)$ Yang-Mills theory also has no bulk phase transition in
$(2+1)$-D and $(3+1)$-D. 

\section{Conjecture}
Our numerical results show that $(2+1)$-D $Sp(2)$ Yang-Mills theory has a second
order deconfinement transition in the $2$-D Ising universality
class. However, $(3+1)$-D $Sp(2)$, $(2+1)$-D and $(3+1)$-D $Sp(3)$
Yang-Mills theories deconfine with a non-universal first order
transition. Hence, despite the fact that a universality class is available,
Yang-Mills theory can have a non-universal first order deconfinement
transition. A non-trivial center plays no role in determining the order of
this transition. Instead our $Sp(N)$ and the $SU(N)$ \cite{Luc03} results
indicate that the order of the deconfinement transition is a
dynamical issue related to the size of the gauge group.
We conjecture that the difference in the number of the relevant
degrees of freedom between the confined phase (color singlet
glueballs) and the deconfined phase (gluon plasma) determines 
the order of the deconfinement transition. Thus, we expect that in $(3+1)$-D
only $SU(2)$ Yang-Mills theory has a second order deconfinement
transition; in $(2+1)$-D, due to stronger fluctuations, only $SU(N)$,
$N=2,3,4$, and $Sp(2)$  Yang-Mills theories should have second order
transitions. According to this picture, $E(6)$ and $E(8)$ Yang-Mills
theories should also have a first order transition due to the large
size of the groups: 78 and 133 generators, respectively. 
For Yang-Mills theories with trivial center gauge  groups
$G(2)$, $F(4)$, $E(8)$ there is no compelling reason for a finite
temperature phase transition \cite{Hol03}. Then, although a first
order transition can not be ruled out, we expect a crossover. 

\end{document}